\documentclass{PoS}

\title{Towards the QCD equation of state at the physical point using Wilson fermion}

\ShortTitle{Towards the QCD equation of state at the physical point using Wilson fermion}

\author{\speaker{T. Umeda}\\
        Graduate School of Education, Hiroshima University, Hiroshima 739-8524, Japan\\
E-mail: \email{tumeda@hiroshima-u.ac.jp}}

\author{S. Ejiri\\
        Department of Physics, Niigata University, Niigata 950-2181, Japan}

\author{R. Iwami\\
        Graduate School of Science and Technology, Niigata University, Niigata 950-2181, Japan}

\author{K. Kanaya\\
        Center for Integrated Research in Fundamental Science and Engineering (CiRfSE),\\
        Faculty of Pure and Applied Sciences, University of Tsukuba, Tsukuba, Ibaraki 305-8571, Japan}

\author{(WHOT-QCD Collaboration)}

\abstract{We study the (2+1)-flavor QCD at nonzero temperatures using nonperturbatively improved Wilson quarks of the physical masses by the fixed scale approach. We perform physical point simulations at finite temperatures with the coupling parameters which were adopted by the PACS-CS Collaboration in their studies using the reweighting technique. Zero temperature values are obtained on the PACS-CS configurations which are open to the public on the ILDG/JLDG. Finite temperature configurations are generated with the RHMC algorithm. The lattice sizes are $32^3 \times N_t$ with $N_t=14$, 13, $\cdots$, 4 which correspond to $T \approx 160$--550 MeV. We present results of some basic observables at these temperatures and the status of our calculation of the equation of state.}

\FullConference{The 33rd International Symposium on Lattice Field Theory\\
		14 -18 July 2015\\
		Kobe International Conference Center, Kobe, Japan*}

\begin{document}
\section{Motivations}

Study of the equation of state (EOS) is one of the main subjects in finite-temperature lattice QCD. 
The EOS plays a key role in understanding thermodynamic and hydrodynamic properties the quark gluon plasma in heavy ion collision experiments. 
Recently, the EOS in (2+1)-flavor QCD at the physical point has been evaluated using staggered-type lattice quarks \cite{Borsanyi:2013bia, Bazavov:2014pvz}. 
%%%%%%
However, theoretical basis of the staggered-type fermions such as locality and universality are not well established.
Therefore, it is important to check the results with theoretically sound lattice fermions such as the Wilson-type fermions.

In order to calculate the EOS in QCD using Wilson-type quarks, we have to address the problem of large computational cost. 
As a possible way out, 
we have developed the fixed-scale approach, in which the temperature $T=(N_ta)^{-1}$ is varied by varying a temporal lattice size $N_t$ at a fixed lattice spacing $a$ 
\cite{Umeda:2008bd,Umeda:2012er,Ejiri:2012vy}.
Because the approach can fix the coupling parameters for all temperatures, zero-temperature simulations to renormalize observables at finite temperatures are common. 
Furthermore the requirement to be on a line of constant physics is automatically satisfied in this approach.  
This is in contrast to the case of the conventional fixed-$N_t$ approach, in which a big computational burden to carry out a series of zero-temperature simulations is required to determine the line of constant physics in the coupling parameter space. 

In the fixed-scale approach, the trace anomaly $\epsilon -3p$ is calculated as usual at each temperature, 
where $\epsilon$ and $p$ are the energy density and the pressure, respectively,
To calculate the pressure non-perturbatively, we adopt the `$T$-integration method' \cite{Umeda:2008bd}:
\begin{eqnarray}
\frac{p}{T^4} = \int^{T}_{T_0} dT \, \frac{\epsilon - 3p}{T^5}
\label{eq:Tintegral}
\end{eqnarray}
with $p(T_0) \approx 0$.
In a previous study, we have shown that the method works well to calculate the EOS of (2+1)-flavor QCD using a clover-improved Wilson fermion, whose quark masses are, however, heavier than the physical masses \cite{Umeda:2012er}.

As the next step we extend our study of the EOS to the physical point.
The fixed-scale approach enables us to calculate the EOS borrowing zero-temperature configurations which are public on the ILDG.
We adopt the zero-temperature configurations generated by the PACS-CS Collaboration, who carried out a series of physical point simulations in (2+1)-flavor QCD using the improved Wilson fermion \cite{Aoki:2009ix}.
In the next section we discuss our simulation parameters for (2+1)-flavor QCD with Wilson fermion at zero and finite temperatures. 
Numerical results are shown in Sec.~3. Summary and outlook are given in the last section. 

\section{Lattice setup}
\subsection{Formulation}

Our QCD action $S=S_g+S_q$ is defined by a combination of the RG-improved gauge action $S_g$ and the $O(a)$ improved Wilson quark action $S_q$ as follows:
\begin{eqnarray}
S_g &=& -\beta\left\{ 
\sum_{x,\mu>\nu}c_0W^{1\times 1}_{\mu\nu}(x)
+\sum_{x,\mu,\nu}c_1W^{1\times 2}_{\mu\nu}(x)
\right\},\label{eq:sg}\\
S_q &=& \sum_{f=u,d,s}\sum_{x,y} \bar{q}_x^f D^{(f)}_{x,y}q_y^f, \label{eq:sq}\\
D^{(f)}_{x,y} &=& \delta_{x,y}-\kappa_f
\sum_\mu\{ (1-\gamma_\mu)U_{x,\mu}\delta_{x+\hat{\mu},y}
+(1+\gamma_\mu)U^\dagger_{x-\hat{\mu},\mu}\delta_{x-\hat{\mu},y}
\}
-\delta_{x,y}c_{SW}\kappa_f\sum_{\mu>\nu}\sigma_{\mu\nu}
F_{\mu\nu},
\nonumber
\end{eqnarray}
with $c_1=-0.331$ and $c_0=1-8c_1$. The clover coefficient $c_{SW}$ is nonperturbatively determined in Ref.~\cite{Aoki:2005et} as a function of $\beta$.
The action is the same as that adopted in our previous study on the EOS in (2+1)-flavor QCD with heavier quarks \cite{Umeda:2012er} and in the physical point simulation by the PACS-CS Collaboration \cite{Aoki:2009ix}.

The trace anomaly is given by 
\begin{eqnarray}
\frac{\epsilon-3p}{T^4}&=&\frac{N_t^3}{N_s^3}
\left(
{a\frac{d\beta}{d a}}
\left\langle
\frac{\partial S}{\partial\beta}
\right\rangle_{\!\!{\rm sub}}
+\,{a\frac{d\kappa_{ud}}{d a}}
\left\langle
\frac{\partial S}{\partial \kappa_{ud}}
\right\rangle_{\!\!{\rm sub}}
+\,{a\frac{d\kappa_s}{d a}}
\left\langle
\frac{\partial S}{\partial \kappa_s}
\right\rangle_{\!\!{\rm sub}}
\right) \label{eq:tranom}\\
\left\langle 
\frac{\partial S}{\partial \beta} 
\right\rangle &=&N_s^3N_t\left(
-\left\langle
\sum_{x,\mu>\nu}c_0W^{1\times 1}_{\mu\nu}(x)
+\sum_{x,\mu,\nu}c_1W^{1\times 2}_{\mu\nu}(x)
\right\rangle \right.\nonumber\\
&&\left.
+ \; \sum_f N_f\,\frac{\partial c_{SW}}{\partial \beta}\kappa_f
\left\langle
\sum_{x,\mu>\nu}\mbox{Tr}^{(c,s)}\sigma_{\mu\nu}F_{\mu\nu}
(D^{(f)\,-1})_{x,x}
\right\rangle
\right)\label{eq:dsdb}\\ 
\left\langle 
\frac{\partial S}{\partial \kappa_f} 
\right\rangle &=&N_fN_s^3N_t\left(
\left\langle
\sum_{x,\mu}\mbox{Tr}^{(c,s)}
\{(1-\gamma_\mu)U_{x,\mu}(D^{(f)\,-1})_{x+\hat{\mu},x}
+(1+\gamma_\mu)U^\dagger_{x-\hat{\mu},\mu}
(D^{(f)\,-1})_{x-\hat{\mu},x}
\}
\right\rangle \right.\nonumber\\
&&\left.
+c_{SW}
\left\langle
\sum_{x,\mu>\nu}\mbox{Tr}^{(c,s)}\sigma_{\mu\nu}F_{\mu\nu}
(D^{(f)\,-1})_{x,x}
\right\rangle
\right)\label{eq:dsdk}
\end{eqnarray}
where $\langle\cdots\rangle_{\rm sub}$ means that the corresponding $T=0$ value is subtracted, and $N_f=2$ for degenerate up and down quarks $f=ud$ while $N_f=1$ for the strange quark $f=s$. 
The derivative $\displaystyle{\frac{\partial c_{SW}}{\partial\beta}}$ is obtained from the data of Ref.~\cite{Aoki:2005et}. 
The derivatives $\displaystyle{a \frac{d\beta}{da}}$, $\displaystyle{a \frac{d\kappa_{ud}}{da}}$, and $\displaystyle{a\frac{d\kappa_{s}}{da}}$ are the beta functions defined as the derivatives along the line of constant physics, which have to be determined by a zero-temperature study.
We compute the traces in Eqs.~(\ref{eq:dsdb}) and (\ref{eq:dsdk}) by the random noise method, in which the color and spinor indices are treated separately to suppress statistical error due to finite number of noise vectors.

\subsection{Zero-temperature simulations}

We use the zero-temperature configurations by the PACS-CS Collaboration to calculate zero-temperature observables for the subtraction. 
The configurations are generated at $\beta=1.90$, $\kappa_{ud}=0.137785$, $\kappa_s=0.136600$, and $c_{SW}=1.715$ on the $32^3 \times 64$ lattice. 
Twofold-mass-preconditioned domain-decomposed HMC algorithm for the degenerate up-down quarks, and the UV-filtered PHMC algorithm for the strange quark. The algorithm is made exact with the global metropolis test.

By the reweighting method, the physical point is determined as $\kappa_{ud}=0.13779625$ and $\kappa_s=0.13663375$ at the same $\beta$ and $c_{SW}$. 
The lattice spacing is determined as $a=0.08995(40)$ fm from the results of $m_\pi$, $m_K$, and $m_\Omega$ at the physical point. 
The spatial lattice size of $N_s=32$ corresponds to 3 fm.

At this simulation point, the PACS-CS Collaboration made 80 configurations public on the ILDG/JLDG. % \cite{}. 
They are sampled in a simulation of 2000 MD time units.
Using these configurations and their reweighting factors, we calculate the zero-temperature observables at the physical point. 

\subsection{Finite-temperature simulations}

We generate the finite-temperature gauge configurations just at the physical point determined by the PACS-CS Collaboration,
{\it i.e.} at $\beta=1.90$, $\kappa_{ud}=0.13779625$, $\kappa_s=0.13663375$, and $c_{SW}=1.715$, on the $32^3 \times N_t$ lattices with $N_t=14$, 13, $\cdots$, 5, 4.

For the  finite-temperature simulations we use the Bride++ code \cite{bridge}.
We employ (threefold) mass-preconditioned HMC algorithm for the degenerate up-down quarks, and the RHMC algorithm for the strange quark. Since the code does not employ the even-odd preconditioning, simulations with the odd values of $N_t$ are also performed to increase the number of simulation temperatures. 
The simulations were performed on the BlueGene/Q at KEK.

\begin{figure}[bt]
  \begin{center}
    \includegraphics[width=95mm]{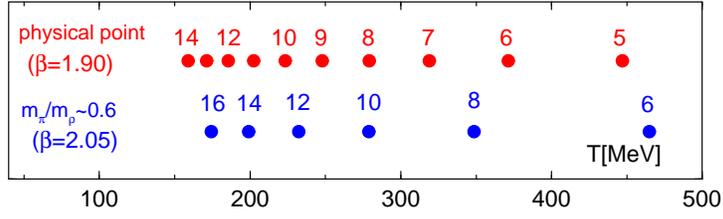}
\caption{
Estimated temperatures for each $N_t$.
The red points are those in this study at the physical point with $\beta=1.90$. 
The blue points are those of our previous study at $m_\pi/m_\rho \approx 0.6$, $\beta=2.05$ \cite{Umeda:2012er}.
   }
    \label{fig:1}
  \end{center}
\end{figure}

From the lattice scale estimated by the CP-PACS Collaboration, our range of $N_t$ corresponds to the range $T \approx 160$--550 MeV.
We show the temperatures of our simulations in Fig.~\ref{fig:1}.
For reference, we also show those of our previous study at the heavier quark mass  ($m_\pi/m_\rho \approx 0.6$) \cite{Umeda:2012er}.
In our previous study, simulations were performed only at even values of $N_t$. 
In the present study, we perform simulations also at odd values of $N_t$.
This enables us to carry out a finer temperature scan even with the smaller value of $\beta$ than the previous study.

\begin{figure}[bt]
  \begin{center}
    \begin{tabular}{ccc}
    \includegraphics[width=70mm]{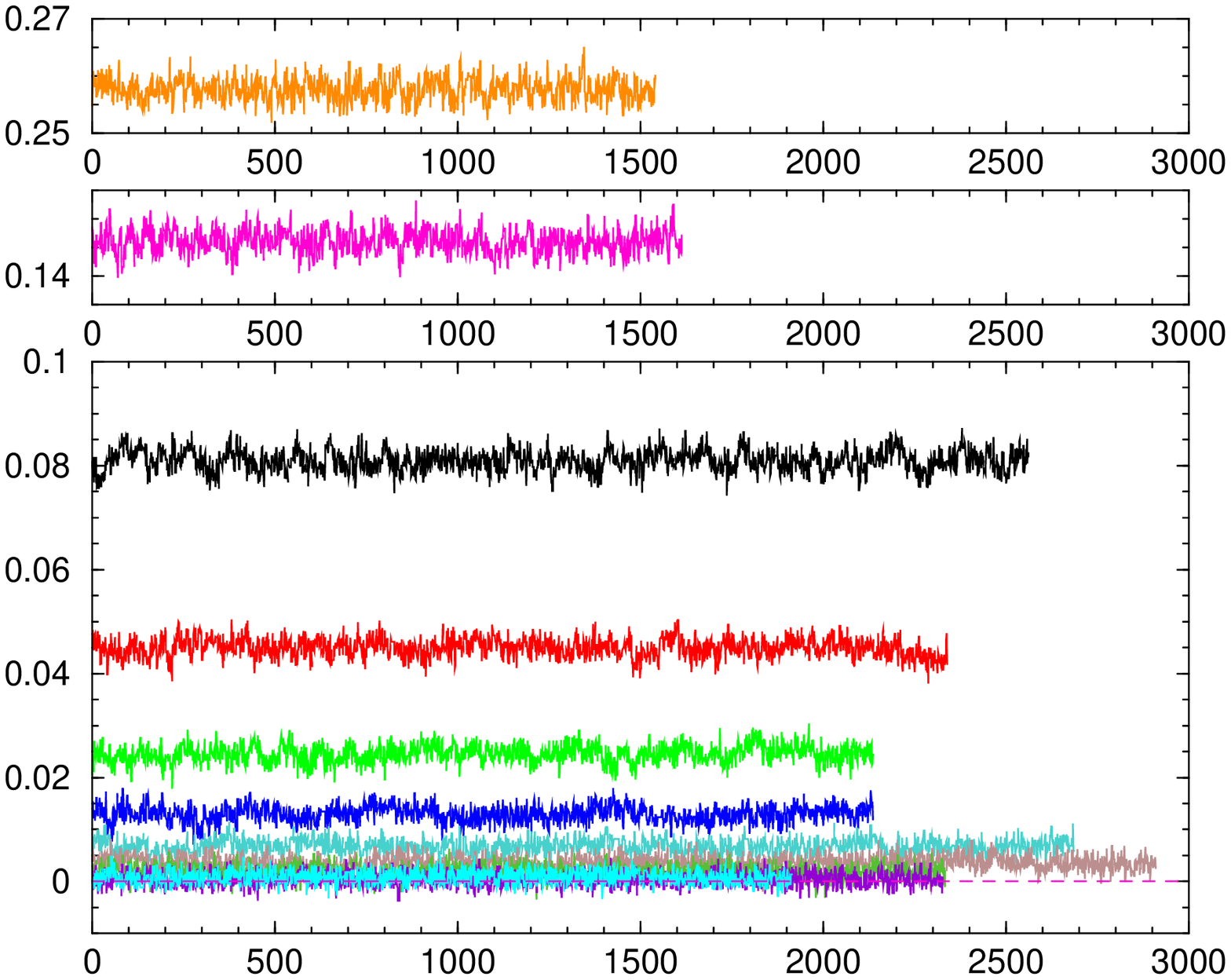}& 
    \includegraphics[width=70mm]{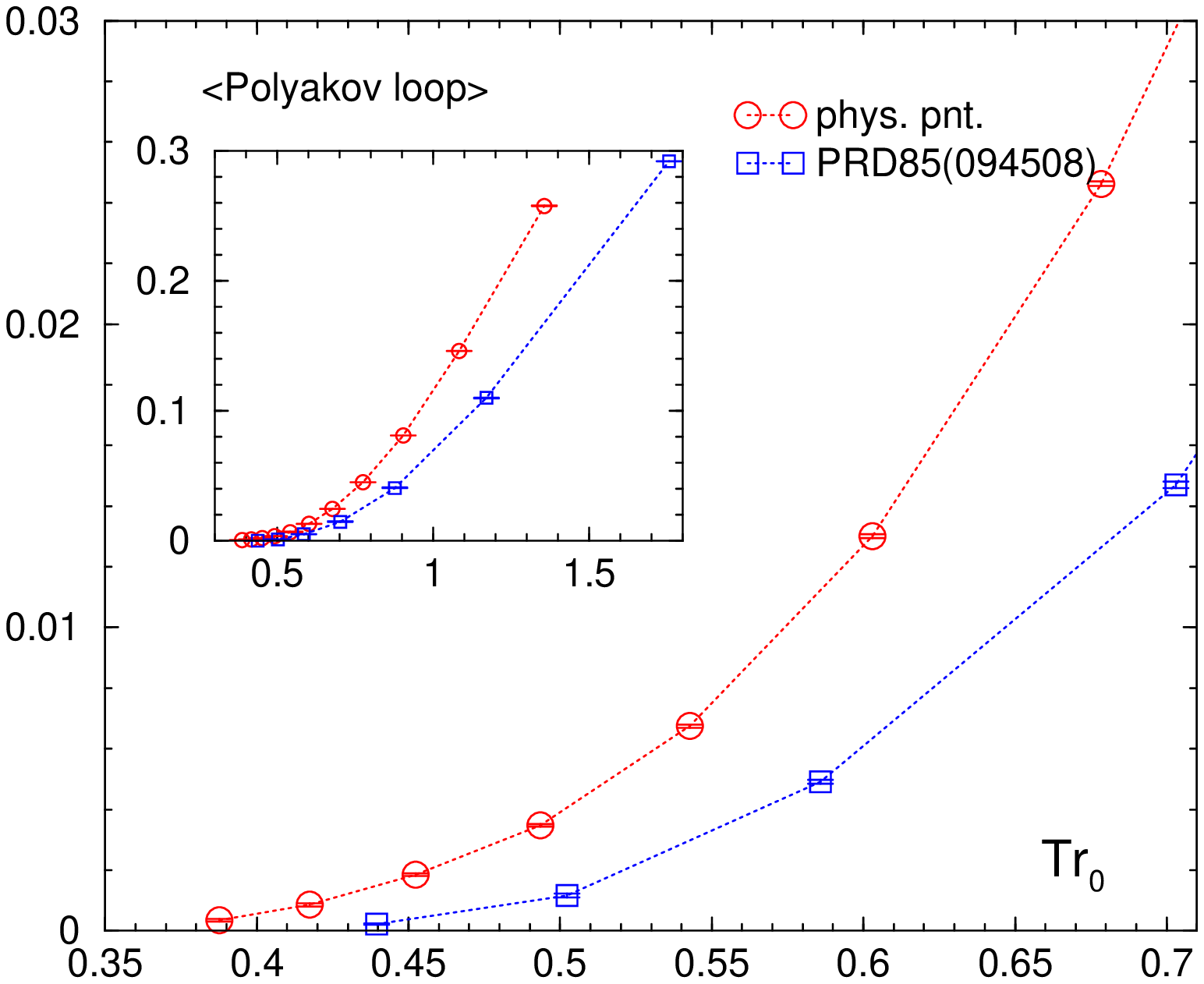} 
    \end{tabular}
\caption{{\em Left}: 
Time history of the Polyakov loop obtained on our finite-temperature lattices. 
In the top and middle panels, the histories of $N_t=4$ and 5 simulations are shown, respectively.
In the panel at the bottom, the histories at $N_t=6$, 7, $\cdots$, 14 are shown in order from the top. 
The horizontal axis is the trajectory length in MD time units.
    {\em Right}: 
The red points are the Polyakov loop expectation values at the physical point as a function of $T r_0$.
The blue points are the results of our previous study at $m_\pi/m_\rho \approx 0.6$.
   }
    \label{fig:2}
  \end{center}
\end{figure}

In the left panel of Fig.~\ref{fig:2}, we show the time histories of the (bare) Polyakov loop measured on our finite-temperature lattices. 
The histories are for $N_t=4$, 5, $\cdots$ from the top to the bottom, respectively. 
Several simulations are still ongoing.

\section{Numerical results at the physical point}

In the right panel of Fig.~\ref{fig:2}, the (bare) Polyakov loop expectation values at the physical point are shown as a function of $T r_0$. 
For the horizontal axis we use $r_0/a=5.427(51)(+81)(-2)$ obtained by the CP-PACS Collaboration \cite{Aoki:2008sm}. 
In this figure, we also plot the Polyakov loop at the heavier quark mass using the result of $r_0/a$ given in Ref.~\cite{Maezawa:2011aa}. 
We see that, when we decrease the quark mass down to the physical point, the Polyakov loop starts to increase from a lower temperature.
This is similar to that observed with staggered-type quarks \cite{Borsanyi:2013bia, Bazavov:2014pvz}.
% Although the Polyakov loop does receive a temperature-dependent renormalization \cite{Gavai:2010qd}, calculation of the renormalization factor is not yet done.

\begin{figure}[bt]
  \begin{center}
    \begin{tabular}{ccc}
    \includegraphics[width=70mm]{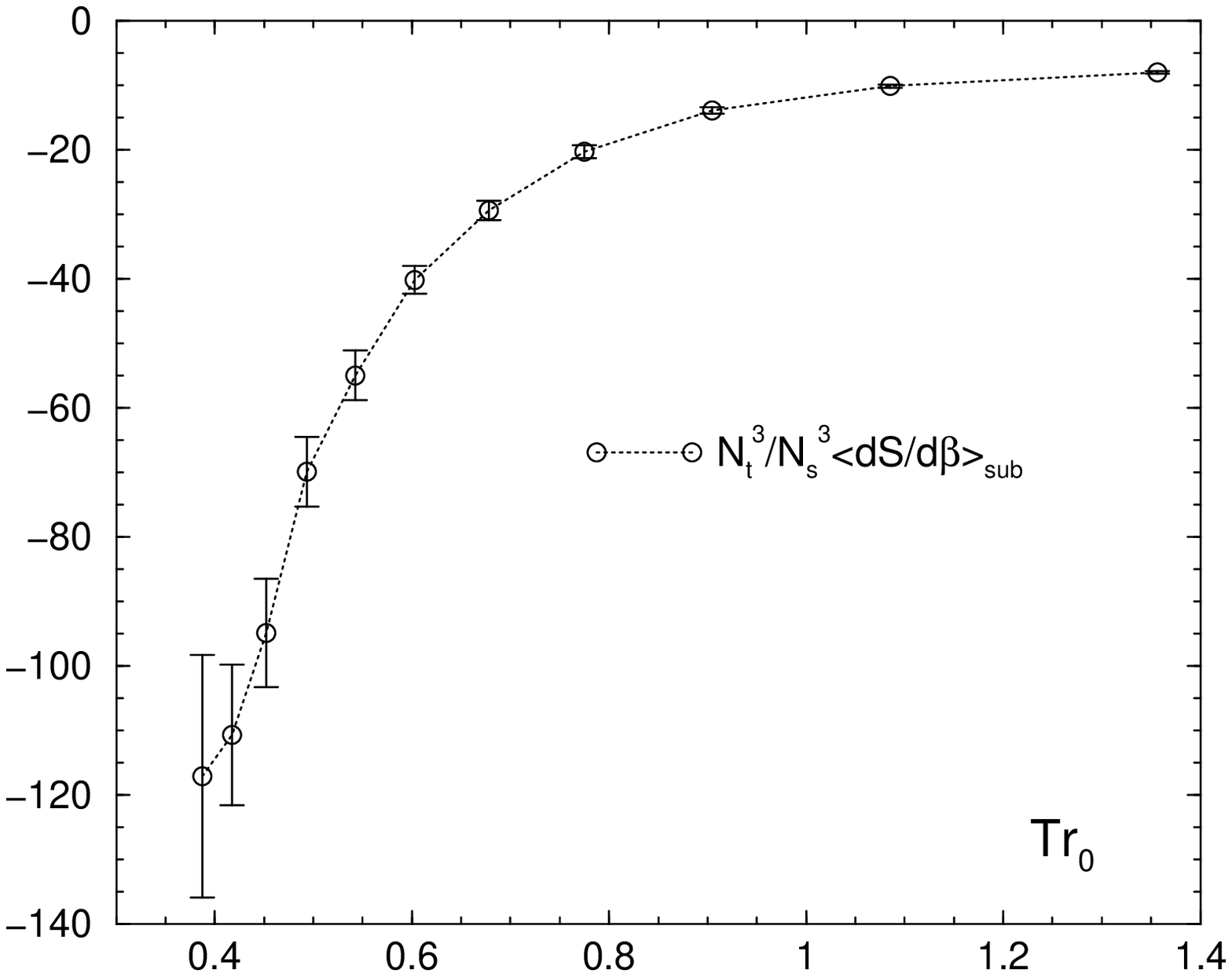}& 
    \includegraphics[width=70mm]{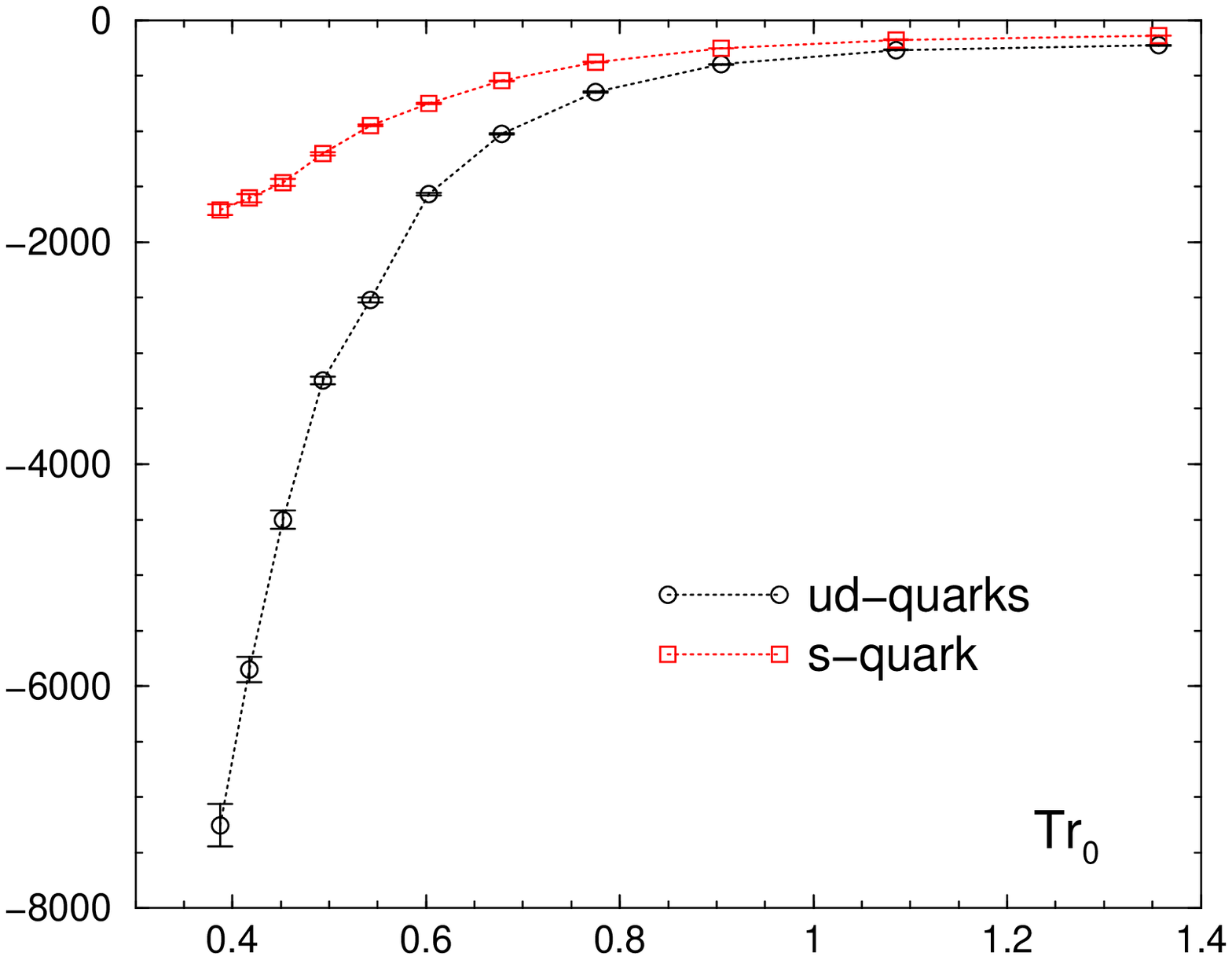} 
    \end{tabular}
\caption{
$(N_t^3/N_s^3) \left\langle \partial S / \partial \beta \right\rangle_{\rm sub}$ (Left) and $(N_t^3/N_s^3) \left\langle \partial S / \partial \kappa_f \right\rangle_{\rm sub}$ (Right) as functions of $T r_0$.
%which appear in (\protect\ref{eq:tranom}) for the trace anomaly.
}
    \label{fig:3}
  \end{center}
\end{figure}

As a step towards calculation of the trace anomaly, we show in Fig.~\ref{fig:3} the results of the $\beta$ and $\kappa$ derivatives of the QCD action with the zero-temperature values subtracted:
$\left\langle \partial S / \partial \beta \right\rangle_{\rm sub}$ and 
$\left\langle \partial S / \partial \kappa_f \right\rangle_{\rm sub}$
defined by (\ref{eq:dsdb}) and (\ref{eq:dsdk}), as functions of $Tr_0$. 
Errors shown in the plots are estimated by combining errors for finite-temperature and zero-temperature parts by the jackknife method with bins of 25 trajectories, using the error propagation formula.
To evaluate the trace anomaly itself according to (\ref{eq:tranom}), we need the values of the beta functions
$a\frac{d\beta}{da}$, $a\frac{d\kappa_{ud}}{da}$, and $a\frac{d\kappa_s}{da}$.
We are still on the way to evaluate them.

%\subsection{Beta-function at the physical point}

\section{Summary and Outlook}

In these proceedings we report the status of our project to calculate the EOS at the physical point using improved Wilson fermion.
The fixed-scale approach enabled us to use the zero-temperature configurations previously generated by the PACS-CS Collaboration at the physical point.
Adopting the coupling parameters of the PACS-CS configurations, we perform finite-temperature simulations just at the physical point on a series of finite-temperature lattices.
To make a finer temperature scan, we simulate also with odd values of $N_t$.

The next step is to evaluate the beta functions at our simulation point. 
In our previous study, we have adopted the direct fit method to evaluate them \cite{Umeda:2012er}. 
Naively, the method requires additional zero-temperature simulations around the physical point, that is quite costly.
Alternatively, we may apply the reweighting method to calculate observables around the physical point. 
Several reweighting factors around the physical point were measured also by the PACS-CS Collaboration in their study of the physical point search.
In order to increase the precision of the beta-functions, furthermore, we may further combine the multi-point reweighting method \cite{Iwami:2015eba} by adding a few simulation points at zero-temperature. 

A completely different approach is to adopt the gradient flow method to compute the EOS through the energy-momentum tensor \cite{Suzuki:2013gza}. 
Because an integration such as (\ref{eq:Tintegral}) is not necessary in this approach, we can avoid statistical noises from the low temperature region. 
On the other hand, perturbative coefficients are assumed, whose validity has to be checked on finite lattices. 
We have just started a study of the EOS with dynamical quarks by the gradient flow method \cite{Itou:2015gxx}. 
Combining different approaches, we are trying to extract reliable results for the EOS with improved Wilson quarks.

\vspace{5mm}
We thank the members of the PACS-CS Collaborations for 
providing us with their physical point configurations and their reweighting factors in (2+1)-flavor QCD,
and other members of the WHOT-QCD Collaboration for valuable discussions.
This work is in part supported by JSPS KAKENHI Grant 
No.\ 26287040, No.\ 26400244, No.\ 26400251, and No.\ 15K05041.
This work is in part supported also by the Large Scale Simulation Program of High Energy Accelerator Research Organization (KEK) No.\ 13/14-21 and No.\ 14/15-23.
Numerical calculations are performed in part using the Bride++ code \cite{bridge}.

\end{document}